# Predicting Intrinsic Antiferromagnetic and Ferroelastic MnF$_4$ monolayer with Controllable Magnetization

Shaowen Xu[a], Fanhao Jia[a], Xuli Cheng[a], and Wei Ren[a*]

[1]*Physics Department, Shanghai Key Laboratory of High Temperature Superconductors, State Key Laboratory of Advanced Special Steel, International Centre of Quantum and Molecular Structures, Shanghai University, Shanghai 200444, China*

[†]renwei@shu.edu.cn

## ABSTRACT

Two-dimensional (2D) multiferroic materials with controllable magnetism have promising prospects in miniaturized quantum device applications, such as high-density data storage and spintronic devices. Here, using first-principles calculations, we propose a coexistence of antiferromagnetism and ferroelasticity in multiferroic MnF$_4$ monolayer. The MnF$_4$ monolayer is found to be an intrinsic wide-gap semiconductor with large spin polarization ~3$\mu_B$/Mn, in which the antiferromagnetic order originates from the cooperation and competition of the direct exchange and super exchange. MnF$_4$ monolayer is also characterized by strongly uniaxial magnetic anisotropic behavior, that can be manipulated by the reversible ferroelastic strain and carrier doping. Remarkably, the carrier doping not only leads to an antiferromagnetic to ferromagnetic phase transformation, bult also could switch the easy magnetization axis between the in-plane and out-of-plane directions. In addition, the Néel temperature was evaluated to be about 140 K from the Monte Carlo simulations based on the Heisenberg model. The combination of antiferromagnetic and ferroelastic properties in MnF$_4$ monolayer provides a promising platform for studying the magnetoelastic effects, and brings about new concepts for next-generation nonvolatile memory and multi-stage storage.

## INTRODUCTION

2D multiferroic materials simultaneously possess two or more ferroic orders[1], typically including ferromagnetic (FM) / antiferromagnetic (AFM), ferroelectric (FE) and ferroelastic (FA) degrees of freedom. The coupling effect of different ferroic properties usually offers an effective way to regulate one ferroicity with another, which is of importance in device applications such as non-volatile memory and magnetic sensors[2-8].

So far, the researches of 2D multiferroics are focused on ferromagnetic and ferroelectric materials [9-12], leaving the AFM based multiferroics rarely reported[13,14]. Fortunately, 2D AFM materials (e.g. $MnPSe_3$[15] and $MnBi_2Te_4$[16]) have been successfully prepared in experiment, which might show some inherent advantages over FM materials in spintronics. The unique zero-net magnetic moment makes the devices insensitive to the external magnetic field, so that they can preserve the signal of magnetoresistance under the continuous reduction in the size of devices[17,18]. On the other hand, AFM materials are more suitable to build high-speed spintronic devices than FM materials because of the higher switching frequencies between different AFM states [19,20]. A few AFM-FA multiferroics have been proposed in theory, for instance, $AgF_2$[20], $VF_4$[21], and FeOOH [22], and FA shows better compatibility with AFM owing to it is neither required the centrosymmetric broken [23]. Therefore, searching and designing new 2D materials with strong AFM-FA coupling will be an important research direction.

Nowadays, various methods have been employed to control the magnetism in the 2D limit [24,25]. One of the most widely used approaches is electrostatic doping, which can effectively tune the spin direction, magnetic anisotropy, and the phase transition temperature. For example, electrostatic doping is able to switch the magnetization axis in CrX (X=P, As) [26] and $CrI_3$ [27] monolayers. Despite rapid development in 2D magnetic materials, numerous layered configurations are unexplored and it is crucial to obtain controllable magnetism for nanoscale spintronic devices.

In this work, we proposed an AFM based multiferroic $MnF_4$ monolayer. We first confirmed the dynamical and thermal stability, and its possibility of experimental exfoliation. $MnF_4$ is intrinsic antiferromagnetic semiconductor with large spin polarization ~$3\mu_B$/Mn. The Néel temperature evaluated from the Monte Carlo simulations based on the Heisenberg model is about 140 K. Additionally, it is also a ferroelastic material with the barrier energy ~160 meV/formula unit (f.u.). More interestingly, hole and electron doping are found to switch the magnetization easy axis in between the in-plane and out-of-plane directions, allowing the effective control of spin injection/detection in 2D structures. Our results are expected to provide a promising 2D multiferroic system for realizing multifunctional devices.

**METHODS**

The first-principles calculations in the frame of the generalized gradient approximation (GGA) proposed by Perdew, Burke, and Ernzerhof (PBE) [28] were performed, as implemented in the Vienna *ab initio* simulation package (VASP) [29]. The energy cutoff was chosen to be 600 eV, and we adopted an additional effective Hubbard $U_{\text{eff}}$ = 4.0 eV for Mn to reduce the delocalization error of 3$d$ electrons [30]. The more accurate exchange functionals optB88-vdW proposed by Dion *et al*. [31] is a non-local correlation functional that approximately accounts for dispersion interactions. The Γ centered k-grids were adopted 8×6×1, and we fully relaxed the atomic positions until the maximum force on each atom was less than $10^{-4}$ eV/Å. To solve the well-known problem of underestimating the band gap of PBE, the screened hybrid functional HSE06 [32] was applied to calculate the band gaps. We choose the vacuum layer more than 20 Å to avoid the interaction between periodic images. The spin-orbit coupling (SOC) was also performed to take the orientation dependence of spin and orbital magnetic moment into account. The phonon dispersion was calculated by density-functional perturbation theory (DFPT) as implemented in the PHONOPY package [33,34], which is widely demonstrated to be valid in 2D systems. The carrier doping was done by changing the total number of electrons of system, and a compensating jellium background of opposite charge was added to maintain charge neutrality.

## RESULTS AND DISCUSSIONS

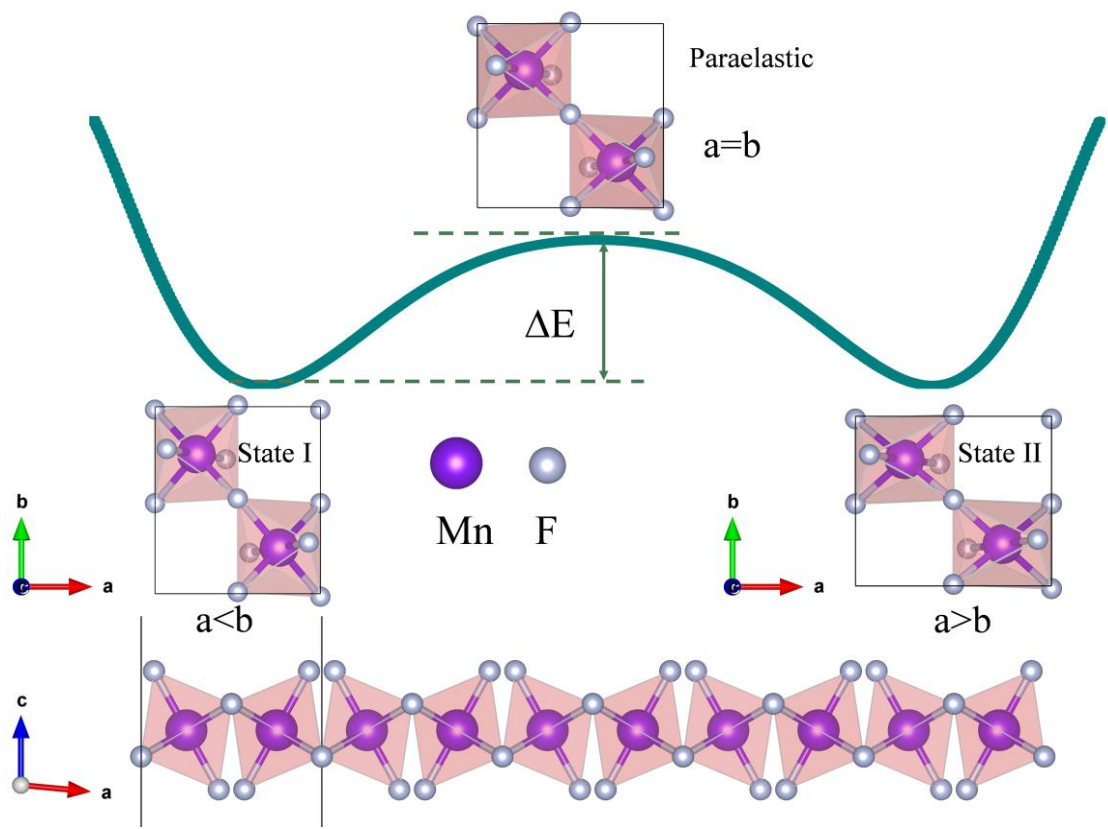

Figure 1 The top and side views of the MnF$_4$ monolayer. The rectangle or square box indicates the unit cell. Schematic diagram of ferroelastic switching between the two different ferroelastic states of the MnF$_4$ monolayer.

Bulk manganese tetrafluoride (MnF$_4$) belongs to the family of transition metal halides, which was predicted to crystallize in a layered monoclinic structure (space group P2$_1$/c)[35]. Due to the van der Waals interaction, MnF$_4$ monolayer is possible to be exfoliated from the layered bulk. Hence, we first verify the stabilities of MnF$_4$ monolayer. There is no appreciable imaginary mode seen in the phonon dispersion (Fig.S1(a)), indicating the dynamtic stability of MnF$_4$ monolayer. Next, we check the thermodynamic stabilities by *ab initio* molecular dynamics (AIMD) simulations at 300K, and observed the 2D planar networks and geometry shapes are well preserved (Fig. S1(b)), which suggests a robust thermal stability of MnF$_4$ monolayer. Additionality, MnF$_4$ also meets the Born criteria: $C_{11}>0$, $C_{66}>0$ and $C_{11}*C_{22} > C_{12}*C_{12}$, indicating it is mechanically stable (see Table SI). The calculated angle-dependent

Young's modulus and Poisson's ratio are presented in Fig. S2. The Young's modulus varies from 30 to 62 N/m, and the Poisson's ratio is ranged from 0.32 to 0.68, showing that MnF$_4$ monolayer has giant mechanical anisotropy.

As shown in Fig.1, the adjacent octahedra tilt around b axis in opposite ways. The Mn ions are caged in the MnF$_6$ octahedron and each Mn atom connects with six neighboring F atoms. The detailed structural parameters are listed in Table I, in which the MnF$_4$ monolayer has the lattice constants of $a = 5.32$ Å, $b = 4.56$ Å. The calculated lengths of in-plane Mn-F bonds are about 1.81 Å (1.88 Å) and the apical Mn-F bond is around 1.926 Å. The angles of F-Mn-F chains are 76.039° (103.97°), and that of Mn-F-Mn are 135.64° (71.80°), as shown in Table I. The unique crystal structure is expected to exhibit ferroelastic properties, with two or more equally stable orientation variants that could switch from one to another without atomic diffusion under external strains[36-38]. Based on the structure of the MnF$_4$ monolayer, the two energy-equivalent ferroelastic (FA) states (state I and state II) and paraelastic (PA) state are displayed in the inset of Fig. 1. The lattice constant $a$ is shorter than $b$ ($a = 4.69$ and $b = 5.37$ Å) in state I, while $a$ is longer than $b$ in state II ($a = 5.37$ and $b = 4.69$ Å). Reversible ferroelastic switching between state I and state II can be achieved under uniaxial tensile strain along the $a$-axis or $b$-axis. The two states are linked by an optimized PA state with $|a| = |b| = 5.06$ Å, such lattice constants are obtained by scanning the square-shaped crystal for the lowest energy. We further confirm the transformation from the PA state to state I or state II is spontaneous, as the phonon dispersion of the PA state indeed has imaginary frequencies. Using this PA state as a reference, the transformation strain matrix $\eta$ can be calculated within Green-Lagrange strain tensor theory[39]:

$$\eta = \frac{1}{2}(\mathbf{J}^T\mathbf{J} - \mathbf{I}) = \frac{1}{2}[(\mathbf{H}_p^{-1})^T\mathbf{H}_f^T\mathbf{H}_f\mathbf{H}_p^{-1} - \mathbf{I}].$$

Here, $\mathbf{H}_p = \begin{pmatrix} 5.06 & 0 \\ 0 & 5.06 \end{pmatrix}$ and $\mathbf{H}_f = \begin{pmatrix} 5.37 & 0 \\ 0 & 4.69 \end{pmatrix}$ represent the lattice constant matrix of PA and FA states, respectively. The $\mathbf{I}$ represents simply the 2 × 2 identity matrix. The $\eta$ follows the form:

$$\eta = \begin{pmatrix} \varepsilon_{xx} & \varepsilon_{xy} \\ \varepsilon_{yx} & \varepsilon_{yy} \end{pmatrix},$$

where $\varepsilon_{xx}$ and $\varepsilon_{yy}$ are the strain along $a$ or $b$-axis. The calculated result is $\eta = \begin{pmatrix} 0.066 & 0 \\ 0 & -0.067 \end{pmatrix}$, suggesting that there is a 6.6% tensile strain along the $a/b$ axis and a 6.7% compressive strain along the $b/a$-axis for the state I / state II.

To evaluate the feasibility of ferroelastic transition, we calculated the energy barrier along the ferroelastic switching pathway. The calculated energy barrier is ~ 160 meV per unit cell, which is comparable to 178 meV per unit cell in CrSCl monolayer[40]. Generally, the moderate energy barrier is thought to beneficial to the applications of ferroelastics. The reversible strain is another important parameter to understand the FA performance, which is defined as (|b|/|a|-1)×100% and calculated to be 14.3%. Thus, MnF$_4$ monolayer is thought to have strong characteristics of ferroelastic switching and obvious structural anisotropic differences. All in all, the obtained switching barrier and reversible strain from our computations suggest promising potential of their experimental realization.

Table I: The lattice constant (Å), bond length (Å) and bond angle (deg.) of MnF$_4$ monolayer calculated by using PBE+$U$ ($U$ = 4eV) method.

| System | Lattice Constant | Bond Length | Bond Angle |
|---|---|---|---|
| MnF$_4$ | $a$ = 5.32 Å | 1.813 Å (Mn-F$_{1/3}$) | 76.0 ∠F$_1$-Mn-F$_2$ |
| | $b$ = 4.56 Å | 1.885 Å (Mn-F$_{2/4}$) | 104.0 ∠F$_2$-Mn-F$_3$ |
| | | 1.926 Å (Mn-F$_{5/6}$) | 180.0 ∠F$_5$-Mn-F$_6$ |
| | | | 135.6 ∠Mn-F$_1$-Mn |
| | | | 135.4 ∠Mn-F$_2$-Mn |

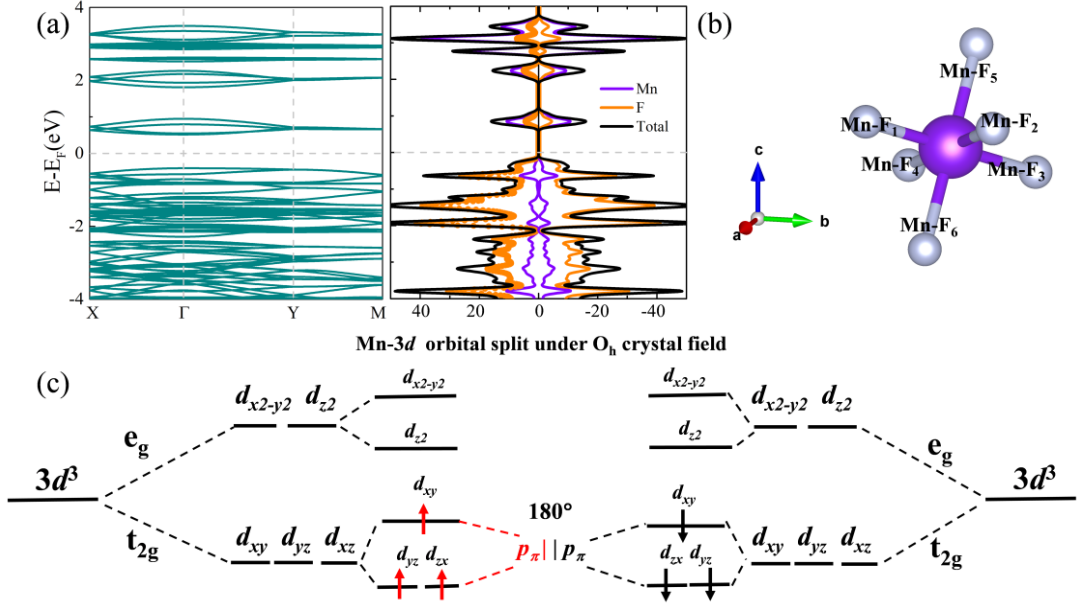

Figure 2 (a) The band structure and partial density of states of MnF$_4$ monolayer. (b) The defined bond length of MnF$_4$ octahedron. (c) The schematic crystal field splitting diagram of Mn 3$d$ orbital.

The MnF$_4$ may possess magnetism owing to that there exists unpaired electron in the Mn$^{4+}$ ions. To determine the magnetic ground state, four 2×2 supercells, namely FM, AFM-1, AFM-2 and AFM-3 configurations are constructed (Fig.S3) in this work. We find the most energetically favorable structure is AFM-2 ordering, which shows the antiferromagnetically alignment of the nearest magnetic moments. To better understand the AFM-2 structure, we present the band structure and density of states (DOS) results in Fig.2(a). The MnF$_4$ monolayer with AFM-2 ordering is found to be semiconductor with a direct band gap around 0.92 eV when using PBE+$U$ method and 2.565 eV for HSE06 method. Both the valence band maximum (VBM) and the conduction band minimum (CBM) are located at the Γ point. The CBM is predominantly contributed by the Mn-$d$ orbitals, while the VBM is derived from F-$p$ orbitals. The spin density shows that spin magnetic moments mainly originate from the Mn atoms ~2.98 $\mu_B$/Mn, which can be understood by the electronic configuration of Mn atom, [Ar]3d$^5$4s$^2$. Due to the bonding to neighboring F atoms, each Mn atom loses 4 $e^-$ and becomes Mn$^{4+}$ with an electronic configuration of [Ar] 3d$^3$. According to the Hund's rule, the remaining 3 $e^-$ on Mn occupy different $d_{xy}$ $d_{yz}$ and $d_{zx}$ orbitals (Fig.2(c)), leading to the magnetic moment of $3\mu_B$, which is close to that obtained from our DFT calculations.

We further illuminated the origin of AFM in MnF$_4$ monolayer by the Goodenough-Kanamori-Anderson rules[41]. The Mn-3$d$ orbitals in the octahedral environments split into double-degenerate $e_g$ and triple-degenerate $t_{2g}$ orbitals (see Fig. 2(c)). Generally, it is believed AFM coupling is preferred for the direct exchange interaction and 180° super exchange interaction including $e_g$-$p_\sigma$/$p_\sigma$-$e_g$ and $t_{2g}$-$p_\pi$/$p_\pi$-$t_{2g}$ chains. In MnF$_4$ monolayer, the nearest neighbor Mn-Mn chain supports the direct-exchange interaction following the $t_{2g}$-$t_{2g}$ linear paths, obeying the Pauli exclusion principle. Since the $3d^3$ is preferred to occupy the low-energy orbital, there is only one possible Mn-F-Mn ($t_{2g}$-$p_\pi$/$p_\pi$-$t_{2g}$) path in MnF$_4$ monolayer. Therefore, our Mn-F-Mn chains of the super-exchange interaction should follow the $t_{2g}$-$p_\pi$/$p_\pi$-$t_{2g}$ path. Based on the above analysis, we attribute the origins of the AFM-2 ordering to the competition between the super-exchange and direct-exchange interactions.

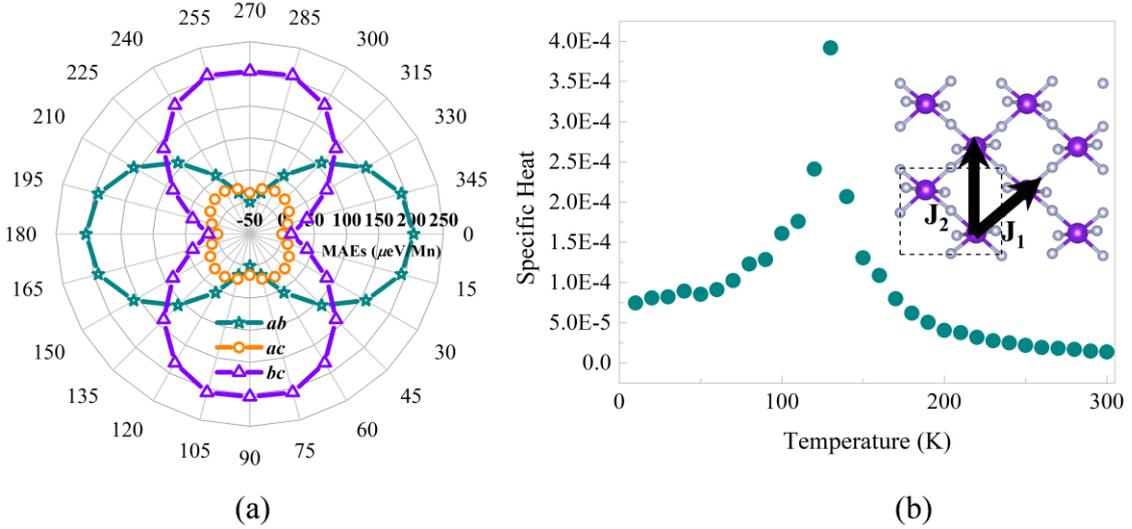

Figure 3 (a) The angle dependent magnetic anisotropy energy (per Mn atom) of MnF$_4$ monolayer. (b) Specific heat as a function of temperature. The $J_1$ and $J_2$ in inset are NN and NNN interaction parameters, respectively.

The spin-orbit coupling (SOC) is also considered to evaluate the magnetic anisotropy energy (MAE) of MnF$_4$ monolayer, and the angle-dependent MAEs are displayed in Fig.3(a). One can see that the magnetic easy axis is along the [010] direction (b axis). The largest MAE is in *ab* plane of ~ 0.204 meV/Mn, suggesting the strong magnetic anisotropy of the MnF$_4$ monolayer. We also employed Monte Carlo (MC) simulations based on the Heisenberg model to simulate the Néel temperature ($T_N$). This strategy has been widely adopted in previous studies. Here, we considered the nearest-neighbour (NN) and next-nearest-neighbour (NNN) exchange interactions. The NN exchange coupling parameter is labeled as $J_1$ and the NNN exchange coupling parameter is $J_2$, as indicated in the insets of Fig. 3(b). The Hamiltonian is written as the following form[42]:

$$H = -2J_1 \sum_{\langle ij \rangle} S_i \cdot S_j - 2J_2 \sum_{\langle\langle ij \rangle\rangle} S_i \cdot S_j - D \sum_i |S_i^z|^2$$

**S** is the spin magnetic moment of the Mn atom and D is the single-ion magnetic anisotropy energy of MnF$_4$ monolayer. A large (50×50) supercell containing 2500 magnetic moments in our MC simulations is used. The computed magnetic susceptibility with respect to temperature (Fig. 3b) shows the $T_N$ values to be 140 K, which well exceeds the liquid nitrogen temperature. Therefore, it indicates MnF$_4$ monolayer might be an intriguing candidate for AFM spintronic applications.

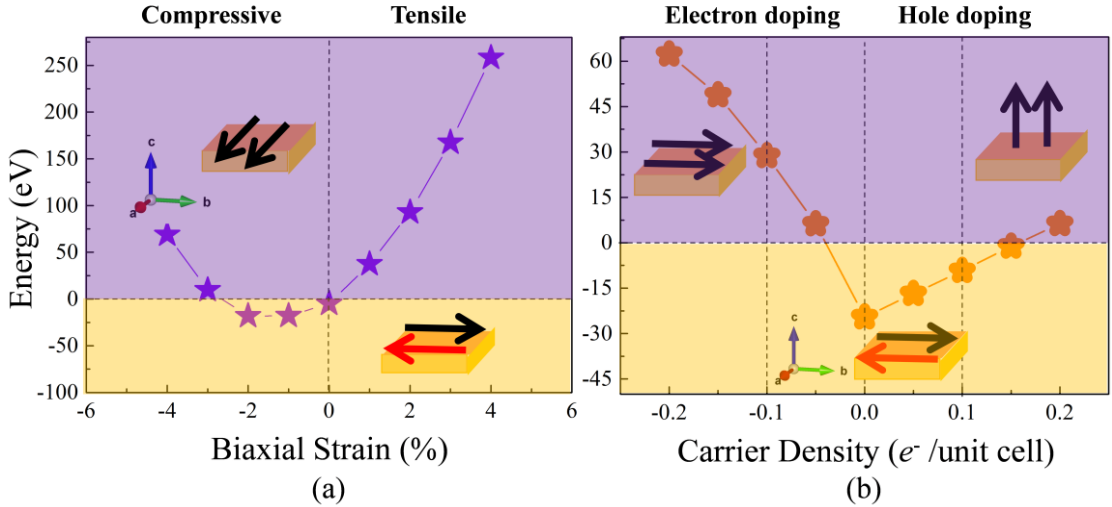

Figure 4 Energy difference of the AFM and FM states as a function of (a) biaxial strain and (b) carrier concentration. The insets represent the easy magnetic axis. The red and black arrows represent the directions of spin up and spin down, respectively. The purple and yellow area represent the FM and AFM states, respectively.

Generally, the external stimuli, such as strain and carrier doping, could effectively tune the electronic and magnetic properties of 2D magnetic materials [7,27,43-45]. We displayed the energy difference between AFM and FM states as a function of biaxial strain in Fig.4(a). Interestingly, the biaxial strain can induce the AFM-FM phase transition accompanying the change of magnetic easy axis. In particular, under the compressive strains in the range of -3% to 0, the MnF$_4$ monolayer prefers to have AFM ground state with the magnetic easy direction along the *b*-axis. Otherwise, the FM states are more energetically favorable, which have the magnetic easy axis along the [100] direction (*a*-axis). Another potential control strategy is through charge carrier doping, which can be realized by applying a negative gate voltage in experiment to inject electrons into the system and produce an electron doping effect, or by adsorption or encapsulation of nucleophilic organic molecules[46]. Remarkably, our results show that the magnetic ground states and the direction of magnetization easy axis can be significantly tuned by carrier doping in MnF$_4$ monolayer. The energy difference between AFM and FM as a function of doping density is shown in Fig.4(b). Under the electron doping, the energy differences between AFM and FM increase with an increase of carrier doping. When the electron doping concentration is larger than 0.05, the MnF$_4$ monolayer changes from AFM to FM and the magnetization easy axis prefers to align along *b*-axis. In contrast, for hole doping larger than 0.15, the MnF$_4$ monolayer can be transformed to FM ground state accompanying the magnetization easy axis from in-plane to out-of-plane orientation. It is worth mentioning that previous experiments have achieved a carrier concentration of $10^{13}$-$10^{14}$ cm$^{-2}$ in 2D systems[26,47] in literature, therefore, the carrier doping is an effective approach to control the MAE in MnF$_4$ monolayer. The charge carrier-tuned magnetic easy axis and magnetic ground states in MnF$_4$ thus provides a promising platform to realize 2D field-effect transistors.

Moreover, the in-plane magnetic easy axis is strongly coupled to the lattice vectors, making it possible to tune the spin orientation via strain. The magnetic easy axis will rotate 90° when state I is changed to state II, with the spin orientation also experiencing a 90° rotation during the ferroelastic transition. Such a functional feature is expected to offer new opportunity for the controllable design of flexible spintronic devices.

## CONCLUSIONS

In summary, we demonstrate that antiferromagnetic and ferroelastic orderings simultaneously exist in MnF$_4$ monolayer. The unique crystal structure of MnF$_4$ monolayer leads to spontaneous ferroelastic bistable states with a moderate activation barrier of 160 meV per formula unit. Mn ions with spin magnetic moments ~3$\mu_B$/Mn are aligned in an antiferromagnetic configuration with the easy magnetization axis along the longer lattice vector direction. Both the spin and easy magnetization axis directions can be regulated by finetuning the carrier doping and biaxial strains. The Néel temperature is predicted to be about 140 K from the Monte Carlo simulations based on the Heisenberg model. Our calculations suggest that the MnF$_4$ system provides a new route for building flexible spintronic devices and we hope our results will stimulate relevant experiments to realize it in the laboratory.

## ACKNOWLEDGMENTS

This work was supported by the National Natural Science Foundation of China (Nos. 51672171, 51861145315 and 51911530124), Shanghai Municipal Science and Technology Commission Program (No.19010500500), State Key Laboratory of Solidification Processing in NWPU (SKLSP201703), Austrian Research Promotion Agency (FFG, Grant No. 870024, project acronym MagnifiSens), and Independent Research Project of State Key Laboratory of Advanced Special Steel and Shanghai Key Laboratory of Advanced Ferrometallurgy at Shanghai University. F.J. is grateful for the support from the China Scholarship Council (CSC).